\documentstyle[12pt]{article}
\begin{document}
\begin{center}
\Large{\bf On the Weak Field Approximation of }\\{\bf Brans-Dicke Theory of Gravity}
\end{center}
\begin{center}
\vspace*{1.5cm}
A. Barros \\ 
{\footnotesize \rm Departamento de F\'\i sica, Universidade Federal de Roraima, \\ 
69310-270, Boa Vista, RR - Brazil.}\\
and\\
C. Romero\footnote{ e-mail: cromero@fisica.ufpb.br }\\
{\footnotesize \rm Departamento de F\'\i sica, Universidade Federal da Para\'\i ba, \\ 
Caixa Postal 5008, 58059-970, Jo\~ao Pessoa, PB - Brazil.}
\end{center}
\vspace*{1.5cm}
\begin{center} {\bf Abstract}
\end{center}

{\footnotesize \rm It is shown that in the weak field approximation solutions of Brans-Dicke equations are simply related to the solutions General Relativity equations for the same matter distribution. A simple method is developed which permits to obtain Brans-Dicke solutions from Einstein solutions when both theories are considered in their linearized forms. To illustrate the method some examples found in the literature are discussed.}

\newpage
$         $

It is a well known fact that most of the mathematical difficulties of General Relativity theory lies in the high non-linearity of the Einstein field equations. On physical grounds this non-linearity means that the gravitational field interacts with itself, and the field contributes to its own source. However, under the special circumstance when the gravitational field is weak one can linearize the field equations thereby ignoring this feedback effect. Such procedure, which leads to a great mathematical simplification of the gravitational field equations, has always found a wide range of applications over the years. In particular, this scheme, often referred to as the weak field approximation has led to the theoretical discovery of ``gravitational waves'', i.e., perturbations of the metric field which satisty the same equations as electromagnetic waves \cite{1}. Other applications of the weak field approximation approach include the spin-$2$ theory of gravity in flat space-time, mostly in connection with attempts to quantize the gravitational field \cite{2}.

Certainly, the weak field approximation technique is not restricted to General Relativity. It has been applied to Brans-Dicke theory of gravity, another metric theory which also makes use of a highly non-linear set of field equations \cite{3}.

In this paper we investigate how solutions of linearized Einstein equations are related to solutions of linearized Brans-Dicke equations when both correspond to the same matter distribution.


To begin with let us recall that in the weak field approximation of General Relativity we assume that the space-time metric tensor deviates only slightly from the flat space-time metric tensor. To put it more precisely we write 

\begin{eqnarray}
\label{2.1}
g_{\mu \nu} = \eta_{\mu \nu} + h_{\mu \nu}\qquad , 
\end{eqnarray}
where $\eta_{\mu \nu}$ denotes Minkowski metric tensor and 
$h_{\mu \nu}$ is to be considered a small perturbation term. The linearized equations are obtained from direct substitution of (\ref{2.1}) into the Einstein's equations keeping only first-order terms in $h_{\mu \nu}$.

On the other hand, Brans-Dicke field equations are given by

\begin{eqnarray}
\label{2.2}
G_{\mu \nu} = {8\pi\over \phi}T_{\mu \nu} + {\omega\over \phi^2}(\phi_{,\mu}\phi_{,\nu}-{1\over 2}g_{\mu \nu}\phi_{,\alpha}\phi^{,\alpha})+{1\over \phi}(\phi_{,\mu;\nu}-g_{\mu \nu}\Box \phi) \qquad ,  
\end{eqnarray}

\begin{eqnarray}
\label{2.3}
\Box \phi ={8\pi T\over 2\omega +3}\qquad ,
\end{eqnarray}
where $\phi$ is a scalar field, $\omega$ is a dimensionless coupling constant and $T$ denotes the trace of the energy-momentum tensor $T_{\mu \nu}.$
Although (\ref{2.2}) and (\ref{2.3}) represent the more usual or standard form of Brans-Dicke equations we are going to consider equivalently the so-called \emph{Einstein representation} \cite{4} given by

\begin{eqnarray}
\label{2.4}
\bar G_{\mu \nu} = 8\pi G_0 \left[\bar T_{\mu \nu} + {2\omega +3\over 16\pi G_0 \phi^2}(\phi_{,\mu}\phi_{,\nu}-{1\over 2}\bar g_{\mu \nu}\phi_{,\alpha}\phi^{,\alpha})\right]\qquad ,  
\end{eqnarray}

\begin{eqnarray}
\label{2.5}
\stackrel{-}{\Box}\ln(G_0 \phi) ={8\pi G_0\over 2\omega +3}\bar T\qquad ,
\end{eqnarray}
which are obtained from (\ref{2.2}) and (\ref{2.3}) by doing the transformation

\begin{eqnarray}
\label{2.6}
\bar g_{\mu \nu}=G_0\phi g_{\mu \nu}\qquad ,
\end{eqnarray}

\begin{eqnarray}
\label{2.7}
\bar T_{\mu \nu}=G_0^{-1}\phi^{-1} T_{\mu \nu}\qquad ,
\end{eqnarray}
where $G_0$ is an arbitrary constant and the bar in $\bar G_{\mu \nu}$, 
$\stackrel{-}{\Box}$ and $\bar T$ just means that these quantities are now calculated using the unphysical metric $\bar g_{\mu \nu}$.

In the weak field approximation of Brans-Dicke theory in addition to (\ref{2.1}) we must also assume that 

\begin{eqnarray}
\label{2.8}
\phi = \phi_0 + \epsilon \qquad ,
\end{eqnarray}
where $\epsilon = \epsilon(x)$ is a first-order term in the energy density and $\left|{\epsilon \over \phi_0} \right| \ll 1$.

Taking into account $(8)$ and setting $G_0=\frac {1}{\phi_{0}}$ the transformation equations (\ref{2.6}) and (\ref{2.7}) become

\begin{eqnarray}
\label{2.9}
\bar g_{\mu \nu}=\eta_{\mu \nu}+\bar h_{\mu \nu} \qquad ,
\end{eqnarray}

\begin{eqnarray}
\label{2.10}
\bar T_{\mu \nu}=(1-\epsilon G_0)T_{\mu \nu}=T_{\mu \nu} \qquad ,
\end{eqnarray}
where 

\begin{eqnarray}
\label{2.11}
\bar h_{\mu \nu}= h_{\mu \nu}+\epsilon G_0\eta_{\mu \nu} \qquad ,
\end{eqnarray}
and only first-order terms in the mass density have been kept.

Now, substituting (\ref{2.8}) in the field equations (\ref{2.4}) and having in view (\ref{2.9}) and  (\ref{2.10}) we get

\begin{eqnarray}
\label{2.12}
\bar G_{\mu \nu}=8\pi G_0 T_{\mu \nu}. 
\end{eqnarray} 
On the other hand, the scalar field equation (\ref{2.3}) becomes

\begin{eqnarray}
\label{2.13}
\Box \epsilon = {8\pi T\over 2\omega +3}. 
\end{eqnarray} 

It turns out then that the equations (\ref{2.12}) are formally identical to the field equations of General Relativity with $G_0$ replacing the Newtonian gravitational constant $G$. Therefore, if $\bar g_{\mu \nu}(G,x)$ is a known solution of the Einstein equations in the weak field approximation for a given $T_{\mu \nu}$, then the Brans-Dicke solution corresponding to the same $T_{\mu \nu}$ will be given in the weak field approximation just by taking the inverse of equation (\ref{2.6}), i.e., 

\begin{eqnarray}
\label{2.14}
g_{\mu \nu}(x)=G_0^{-1}\phi^{-1}\bar g_{\mu \nu}(G_0,x)=[1-\epsilon (x)G_0]\bar g_{\mu \nu}(G_0,x) \qquad ,  
\end{eqnarray} 
or, equivalently,

\begin{eqnarray}
\label{2.15}
h_{\mu \nu}(x)=\bar h_{\mu \nu}(G_0 ,x)-\epsilon (x)G_0\eta_{\mu \nu}. 
\end{eqnarray}

Thus, we conclude that the general problem of finding solutions of Brans-Dicke equations of gravity in the weak field approximation may be reduced to solving Einstein field equations for the same matter distribution.

It should be noted that the Einstein tensor $\bar G_{\mu \nu}$ which appears in the left hand side of (\ref{2.12}) must be calculated in the weak field approximation, i.e., taking $\bar g_{\mu \nu}$ as given by (\ref{2.9}).  

As to the function $\epsilon(x)$, which appears in the conformal factor of the metric $\bar g_{\mu \nu}(G_0,x)$, it may be calculated directly from (\ref{2.13}) and will be given as a retarded integral in the form

\begin{eqnarray}
\label{2.16}
\epsilon(x) =\frac{2}{2\omega +3} \int\frac{T(t-|\vec{x}-\vec{x}'|,\vec{x}')}{|\vec{x}-\vec{x}'|}d^3x' \qquad ,
\end{eqnarray}
being $\eta_{\mu\nu}=$diag$(1,-1,-1,-1)$ in equation (\ref{2.13}).
Let us conclude this paragraph with a remark concerning the constant $\phi_0$, which comes out in the weak field approximation. Actually, in order that Brans-Dicke theory possess a Newtonian limit this constant must be related to the Newtonian gravitational constant $G$ by setting \cite{3}

\begin{eqnarray}
\label{2.17}
{1\over \phi_0}=\left({2\omega +3\over 2\omega +4}\right)G.
\end{eqnarray}
Thus we have

\begin{eqnarray}
\label{2.18}
G_0=\left({2\omega +3\over 2\omega +4}\right)G.
\end{eqnarray}


We shall go through some known solutions of Brans-Dicke equations in the weak field approximation and show how they could be directly obtained
with the help of the method just outlined.

Let us start with the problem of finding out the space-time and the scalar field generated by a static point of mass $M$ in the weak field approximation of Brans-Dicke theory of gravity. The energy-momentum tensor which corresponds to such a configuration is given by

\begin{eqnarray}
\label{3.19}
T_{\nu}^{\mu}=\hbox{diag}(M\delta (\vec{r}),0,0,0).
\end{eqnarray}
Now, it is well known that the solution of the equivalent problem in the weak field approximation of General Relativity expressed in isotropic coordinates is given by the line element \cite{5}

\begin{eqnarray}
\label{3.20}
ds^{2}=\left(1-\frac{2MG}{r}\right)dt^{2}-\left(1+\frac{2MG}{r}\right)
\left[dr^{2}+r^{2}(d\theta ^{2}+\sin^{2}\theta d\varphi^{2})\right].
\end{eqnarray} 

Following the previous reasoning all we have to do is to solve equation (\ref{2.13}), which taking into account (\ref{3.19}), reduces to

\begin{eqnarray}
\label{3.21}
\nabla ^{2}\epsilon =-\frac{8\pi M}{2\omega +3}\delta(\vec{r}).
\end{eqnarray}
The solution of (\ref{3.21}) is easily found:

\begin{eqnarray}
\label{3.22}
\epsilon(r)=\frac{2M}{(2\omega +3)r}.
\end{eqnarray}
Thus, in the weak field approximation of Brans-Dicke theory the space-time generated by a static point mass according to (\ref{2.14}) is given by

\begin{eqnarray}
\label{3.23}
ds^{2}=\left(1-\frac{2MG_0}{(2\omega +3)r}\right)\left[\left(1-\frac{2MG_0}{r}\right)dt^{2}-\left(1+\frac{2MG_0}{r}\right)d\Omega ^{2}\right] \qquad ,
\end{eqnarray} 
with $d\Omega ^{2}=dr^{2}+r^{2}(d\theta ^{2}+\sin^{2}\theta d^{2}\varphi)$.
In this way either from (\ref{3.23}) (neglecting second-order terms in the mass density) or directly from (\ref{2.15}) we have:

\begin{eqnarray}
\label{3.24}
h_{00}=-\frac{2MG_0}{r}\left[\frac{2\omega +4}{2\omega +3}\right]=-\frac{2MG}{r} \qquad ,
\end{eqnarray}

\begin{eqnarray}
\label{3.25}
h_{ii}=-\frac{2MG_0}{r}\left[\frac{2\omega +2}{2\omega +3}\right]=-\frac{2MG}{r}\left(\frac{\omega +1}{\omega +2}\right) \qquad ,
\end{eqnarray}
(with $i=1, 2, 3$) which coincides with the known result obtained by Brans and Dicke \cite{3}.

As a second example let us consider the line element which describes the space-time generated by a static string the energy-momentum tensor of which is given by 

\begin{eqnarray}
T^{\mu}_{\nu}=\delta (x)\delta(y)\hbox{diag}(\mu ,0,0,-p)\qquad , \nonumber
\end{eqnarray}
where $\mu $ is the linear energy density and $p$ is the pressure in the $z$ direction. The solution of this problem in the context of General Relativity was first worked out by Vilenkin \cite{6}. Using the weak field approximation Vilenkin solved the field equations and obtained in cartesian coordinates

\begin{eqnarray}
\label{3.26}
\bar h_{00}=\bar h_{33}=4G(\mu +p)\ln(\rho/\rho_0)\qquad ,
\end{eqnarray}

\begin{eqnarray}
\label{3.27}
\bar h_{11}=\bar h_{22}=4G(\mu -p)\ln(\rho/\rho_0)\qquad ,
\end{eqnarray}
where $\rho=(x^{2}+y^{2})^{1/2}$ and $\rho_{0}$ is a constant. Again, using the weak field approximation to approach this problem in Brans-Dicke theory we have to solve equation (\ref{2.13}) which takes the form:

\begin{eqnarray}
\label{3.28}
\nabla ^{2}\epsilon=-\frac{8\pi}{2\omega +3}(\mu -p )\delta (x)\delta (y) \hspace{0.5cm},
\end{eqnarray}
whose solution is readily found to be

\begin{eqnarray}
\label{3.29}
\epsilon=-\frac{4(\mu -p)}{2\omega +3} \ln\frac{\rho}{\rho_0}.
\end{eqnarray}

Then, from (\ref{2.14}) it follows that the sought-for line element, which describes the space-time generated by the string in Brans-Dicke theory, is given 
by

\begin{eqnarray}
\label{3.30}
ds^{2}&=&\biggr[1+\frac{4(\mu -p)G_{0}}{2\omega +3}\ln{\frac{\rho}{\rho_0}}
\biggr] \biggr[\left(1+4(\mu +p)G_{0}\ln\frac{\rho}{\rho_0}\right)dt^{2} 
\nonumber \\ \nonumber \\ 
& & -\left(1-4(\mu -p)G_{0}\ln\frac{\rho}{\rho_0}\right)(dx^{2}+dy^{2})
 \nonumber \\ \nonumber \\ 
& &- \left(1-4(\mu +p)G_{0}\ln\frac{\rho}{\rho_0}\right)dz^{2} \biggr].
\end{eqnarray}
Particularly, for a vacuum string, $p=-\mu$, and turning to cylindrical coordinates (\ref{3.30}) reduces to 
\begin{eqnarray}
\label{3.31}
ds^{2}=\left[1+\frac{8\mu G_{0}}{2\omega +3} \ln\frac{\rho}{\rho_0}\right]
\left[dt^{2}-dz^{2}-\left(1-8\mu G_{0}\ln\frac{\rho}{\rho_0}\right)
(d\rho^{2}+\rho^{2}d\theta^{2})\right].
\end{eqnarray}
Finally, introducing a new coordinate $\rho^{\prime}$ by the transformation $\rho=\rho_{0}\left(\frac{\rho^{\prime}}{a}\right)^{b}$, where $a=\rho_{0}(1-8\mu G_{0})^{-1/2}$ and $b=(1-4\mu G_{0})^{-1}$, and neglecting second-order terms in $\mu G_{o}$ we arrive at

\begin{eqnarray}
\label{3.32}
ds^{2}=\left[1+\frac{8\mu G_{0}}{2\omega +3} \ln\frac{\rho^{\prime}}{\rho_0}\right]
[dt^{2}-dz^{2}-d\rho^{\prime 2}-(1-8\mu G_{0})\rho^{\prime 2}d\theta^{2}]\qquad ,
\end{eqnarray}
which is the result obtained in ref.\cite{7}.

The third example comes from Vilenkin's solution corresponding to the space-time of a static massive plane \cite{6}. In this case, the source of the gravitational field consists of an infinite static plane wall parallel to the $(y,z)$ plane. For a homogeneous energy surface distribution $\sigma$ the energy-momentum tensor $T^{\mu}_{\nu}$ is given by $T^{\mu}_{\nu}=\delta(x)\hbox{diag}(\sigma,0,-p,-p)$, where $p$ is the pressure. In the weak field approximation of the linearized Einstein equations yield the solution

\begin{eqnarray}
ds^{2}&=&[1+4\pi G(\sigma+2p)|x|]dt^{2}-[1-4\pi G(\sigma-2p)|x|]dx^{2} \nonumber \\
\label{3.33}
& &-[1-4\pi G\sigma |x|](dy^{2}+dz^{2}).
\end{eqnarray}

As before, in order to get the corresponding solution in Brans-Dicke theory, we must solve equation (\ref{2.13}) which in this case will be given by

\begin{eqnarray}
\label{3.34}
\Box\epsilon=\frac{8\pi}{2\omega +3}(\sigma-2p)\delta(x).
\end{eqnarray}
Due to planar symmetry $\epsilon=\epsilon(x)$ and (\ref{3.34}) is reduced to

\begin{eqnarray}
\frac{d^{2}\epsilon}{dx^{2}}=-\frac{8\pi}{2\omega +3}(\sigma-2p)\delta(x)\qquad ,
\nonumber
\end{eqnarray}
hence yielding the solution

\begin{eqnarray}
\label{3.35}
\epsilon=-\frac{4\pi}{2\omega +3}(\sigma-2p)|x|.
\end{eqnarray}
Therefore, from (\ref{2.14}) we obtain

\begin{eqnarray}
ds^{2}&=&\biggr[1+\frac{4\pi G_{0}}{2\omega +3}(\sigma-2p)|x|\biggr]
\biggr[(1+4\pi G_{0}(\sigma+2p)|x|)dt^{2} \nonumber \\
\label{3.36}
& &-(1-4\pi G_{0}(\sigma-2p)|x|)
dx^{2}-(1-4\pi G_{0}\sigma |x|)(dy^{2}+dz^{2})\biggr] ,
\end{eqnarray}
which represents the space-time generated by the static massive plane in Brans-Dicke theory. For a vacuum domain wall, $p=-\sigma$ and

\begin{eqnarray}
\label{3.37}
h_{00}=-h_{22}=-h_{33}=-\frac{8\pi G_{0}\sigma\omega |x|}{2\omega+3}\qquad ,
\end{eqnarray}

\begin{eqnarray}
\label{3.38}
h_{11}=\frac{24\pi G_{0}\sigma(\omega+1)|x|}{2\omega+3}\qquad ,
\end{eqnarray}
in according with ref.\cite{7}.

Finally, to give a last example let us consider the space-time generated by a global monopole \cite{8}. This solution which was obtained also under the weak field approximation, albeit originally it was not derived directly from the linearized Einstein equations, is given by

\begin{eqnarray}
\label{3.39}
ds^{2}=dt^{2}-dr^{2}-(1-8\pi\eta^{2}G)r^{2}(d\theta^{2}+\sin^{2}\theta d\varphi^{2})\qquad ,
\end{eqnarray}
where $\eta$ is the energy scale of the symmetry breaking. The energy-momentum
tensor corresponding to the monopole is given by

\begin{eqnarray}
\label{3.40}
T^{\mu}_{\nu}=\hbox{diag}\left(\frac{n^{2}}{r^{2}},\frac{n^{2}}{r^{2}},0,0\right).
\end{eqnarray}
Thus, our task is to solve equation (\ref{2.13}) taking into account (\ref{3.40}). From the fact that the scalar field must be static and spherically
symmetric equation (\ref{2.13}) becomes

\begin{eqnarray}
\label{3.41}
\nabla^{2}\epsilon=-\frac{16\pi\eta^{2}}{(2\omega+3)r^{2}}.
\end{eqnarray}
Then, keeping only linear terms in $\eta^{2}$ (\ref{3.41}) yields the solution

\begin{eqnarray}
\label{3.42}
\epsilon(r)=-\frac{16\pi\eta^{2}}{2\omega+3}\ln\frac{r}{r_{0}}.
\end{eqnarray}
From the prescription given by (\ref{2.14}) we get the following line element

\begin{eqnarray}
\label{3.43}
ds^{2}&=&\biggr[1+\frac{16\pi\eta^{2}}{2\omega+3}G_{0}\ln\frac{r}{r_{0}}\biggr]\biggr[dt^{2}-dr^{2}-(1-8\pi\eta^{2}G_{0})
 \nonumber \\ \nonumber \\
& & \times r^{2}(d\theta^{2}+\sin^{2}
\theta d\varphi^{2})\biggr]\qquad,
\end{eqnarray}
which is the result obtained in ref.\cite{9}.


To conclude we would like to briefly comment on the result expressed by 
equation (\ref{2.14}). Essentially, this equation means that in the weak
field approximation the metric tensor calculated from Brans-Dicke equations
is quasi-conformally related to the metric tensor calculated from Einstein
equations for the same matter configuration. The term quasi-conform here 
should be understood in the sense that in going from the Einstein solution 
$\bar{g}_{\mu\nu}(G,x)$ to the corresponding Brans-Dicke solution $g_{\mu\nu}(x)$ apart from the scale factor $\lambda(x)\equiv1-\epsilon(x)G_{0}$ one must
replace $G$ for $G_{0}$ in $\bar{g}_{\mu\nu}$, i.e., $g_{\mu\nu}(x)=\lambda(x)
\bar{g}_{\mu\nu}(G_{0},x)$.

An immediate physical consequence of that concerns the trajectories of 
light rays. For it is evident that null geodesics in both space-times 
described by $\bar{g}_{\mu\nu}$ and $g_{\mu\nu}$ are closely related: the only change involved is the replacement of the Newtonian gravitational constant $G$ by the new $\omega$-dependent ``effective'' gravitational constant $G_{0}=\frac{2\omega+3}{2\omega+4}G$. For a value of $\omega$ consistent with solar system observations and experiments, say $\omega\sim 500$ \cite{10}, it means that massless particles
 travelling in the space-time described by $g_{\mu\nu}$ would ``feel'' a decrease in the gravitational strength as  $G_{0}\sim 0,999 G$. 

Finally, it is worth mentioning that in the weak field approximation when 
$\omega\rightarrow\infty$ the Brans-Dicke solution goes over the corresponding
 Einstein solution, although this does not always happen in the case of exact solutions \cite{11}. Indeed, to prove this statement just note that when $\omega\rightarrow\infty$ we have, respectively, from (\ref{2.16}) and (\ref{2.18}) 
that $\epsilon(x)\rightarrow 0$ and $G_{0}\rightarrow G$.\\
$        $\\
{\bf Acknowledgements}\\
$      $\\
C. Romero thanks CNPq (Brazil) for financial support.

\end{document}